\documentclass[11pt]{article}
\usepackage[utf8]{inputenc}
\usepackage{amsfonts}
\usepackage{amsmath}
\usepackage{amssymb}
\usepackage{indentfirst}
\usepackage{graphicx}
\usepackage[colorlinks]{hyperref}
\usepackage{cite}
\usepackage{array}
\usepackage{microtype}
\usepackage{soul}
\usepackage[normalem]{ulem}
\usepackage{cancel}
\usepackage{soul}

\numberwithin{equation}{section}
\allowdisplaybreaks

\makeatletter

\begin{document}

\begin{titlepage}
\vspace{3cm}

\baselineskip=24pt

\begin{center}
\textbf{\LARGE{Three-dimensional Maxwellian Carroll gravity theory and the cosmological constant}}
\par\end{center}{\LARGE \par}

\begin{center}
	\vspace{1cm}
	\textbf{Patrick Concha}$^{\ast}$,
	\textbf{Diego Peñafiel}$^{\bullet}$,
    \textbf{Lucrezia Ravera}$^{\star, \ddag}$,
	\textbf{Evelyn Rodríguez}$^{\dag}$,
	\small
	\\[5mm]
    $^{\ast}$\textit{Departamento de Matemática y Física Aplicadas, }\\
	\textit{ Universidad Católica de la Santísima Concepción, }\\
\textit{ Alonso de Ribera 2850, Concepción, Chile.}
	\\[2mm]
	$^{\bullet}$\textit{Facultad de Ciencias, Universidad Arturo Prat, }\\
	\textit{Iquique, Chile.}
	\\[2mm]
	$^{\star}$\textit{DISAT, Politecnico di Torino, }\\
	\textit{ Corso Duca degli Abruzzi 24, 10129 Torino, Italy.}
	\\[3mm]
	$^{\ddag}$\textit{INFN, Sezione di Torino, }\\
	\textit{ Via P. Giuria 1, 10125 Torino, Italy.}
	\\[3mm]
    $^{\dag}$\textit{Departamento de Física, Universidad del Bío-Bío, }\\
	\textit{Avenida Collao 1202, Casilla 5-C, Concepción, Chile}
	 \\[5mm]
	\footnotesize
	\texttt{patrick.concha@ucsc.cl},
	\texttt{dimolina@unap.cl},
    \texttt{lucrezia.ravera@polito.it},
	\texttt{ekrodriguez@ubiobio.cl},
	\par\end{center}
\vskip 26pt
\begin{abstract}

In this work, we present the three-dimensional Maxwell Carroll gravity by considering the ultra-relativistic limit of the Maxwell Chern-Simons gravity theory defined in three spacetime dimensions. We show that an extension of the Maxwellian Carroll symmetry is necessary in order for the invariant tensor of the ultra-relativistic Maxwellian algebra to be non-degenerate. Consequently, we discuss the origin of the aforementioned algebra and theory as a flat limit. We show that the theoretical setup with cosmological constant yielding the extended Maxwellian Carroll Chern-Simons gravity in the vanishing cosmological constant limit is based on an enlarged extended version of the Carroll symmetry. Indeed, the latter exhibits a non-degenerate invariant tensor allowing the proper construction of a Chern-Simons gravity theory which reproduces the extended Maxwellian Carroll gravity in the flat limit.

\end{abstract}
\end{titlepage}\newpage {} 


\section{Introduction}

Symmetries in Physics play a fundamental role in the analysis and understanding of theories. In this perspective, ultra-relativistic (UR) symmetries, also known as Carroll symmetries, have attracted some interest over recent years. The Carroll symmetries arise in the UR limit, $c \rightarrow 0$, being $c$ the speed of light \cite{LevyLeblond, Bacry:1968zf}. The Carroll group introduced by L\'{e}vy-Leblond emerged as the UR contraction of the Poincar\'{e} group, dual to the non-relativistic (NR) contraction (i.e., $c\rightarrow \infty$) leading to the Galilean group.

Models with Carroll symmetries appear, for instance, in the context of high-energy Physics in the study of tachyon condensation \cite{Gibbons:2002tv}, warped conformal field theories \cite{Hofman:2014loa}, and tensionless strings \cite{Bagchi:2013bga, Bagchi:2015nca, Bagchi:2016yyf, Bagchi:2017cte, Bagchi:2018wsn}. Moreover, further works on Carroll algebras in the context of electrodynamics and brane dynamics can be found for instance
in \cite{Duval:2014uoa,Clark:2016qbj,Basu:2018dub}. On the other hand, the exploration of the Carroll limit corresponding to M2- as well as M3-branes propagating over $D=11$ supergravity backgrounds in M-theory has been considered in \cite{Roychowdhury:2019aoi}.
In \cite{Bergshoeff:2015wma}, flat and Anti-de Sitter (AdS) Carroll spaces were investigated, in the bosonic and in the supersymmetric case, studying the symmetries of a particle moving in such spaces. Models of Carrollian gravity have been developed in \cite{Hartong:2015xda, Bergshoeff:2016soe, Bergshoeff:2017btm,Matulich:2019cdo}.
In particular, in \cite{Bergshoeff:2016soe}, the respective NR and UR limits of the spin-3 Chern-Simons (CS) gravity in three spacetime dimensions were presented.
Subsequently, the AdS Carroll CS gravity theory and its supersymmetric extension were discussed in \cite{Matulich:2019cdo} and in \cite{Ravera:2019ize,Ali:2019jjp}, respectively. More recently, the Carrollian version of the Jackiw-Teitelboim gravity was explored in \cite{Grumiller:2020elf,Gomis:2020wxp}.

On the other hand, there has been a growing interest, at the relativistic level, in exploring symmetries in three spacetime dimensions beyond the Poincaré and AdS ones. It is well assumed that a gravitational theory in three spacetime dimensions can be described by a CS action as a gauge theory, providing a useful setup to approach higher-dimensional models. A non-trivial extension of the Poincaré algebra is given by the Maxwell algebra which, in any dimension $D$, is characterized by the commutator
\begin{equation}
    [\tilde{P}_{A},\tilde{P}_{B}] = \tilde{Z}_{AB}\,.
\end{equation}
The Maxwell algebra has been first introduced in four spacetime dimensions to describe a constant Minskowski spacetime in the presence of an electromagnetic background \cite{Schrader:1972zd,Bacry:1970ye,Bacry:1970du,Gomis:2017cmt}. In arbitrary spacetime dimensions, the Maxwell algebra and generalizations have been useful to relate General Relativity with CS and Born-Infeld (BI) gravity theories \cite{Edelstein:2006se,Izaurieta:2009hz,Concha:2013uhq,Concha:2014vka}. In three spacetime dimensions, the Maxwell symmetries and extensions to supersymmetry and higher-spin, have been explored using the CS formalism with diverse physical implications \cite{Cangemi:1992ri,Duval:2008tr,Salgado:2014jka,Hoseinzadeh:2014bla,Caroca:2017izc,Concha:2018zeb,Concha:2018jxx,Concha:2019icz,Chernyavsky:2020fqs,Adami:2020xkm,Caroca:2021bjo}.\footnote{For other relevant applications of the Maxwell algebras, together with supersymmetric extensions of the latter, we refer the interested reader to, e.g., \cite{deAzcarraga:2010sw,Durka:2011nf,deAzcarraga:2012qj,Concha:2014tca,Penafiel:2017wfr,Ravera:2018vra,Concha:2018ywv,Salgado-Rebolledo:2019kft,Concha:2019eip,Kibaroglu:2020tbr,Cebecioglu:2021dqb}.} Interestingly, the inclusion of a cosmological constant to the Maxwell CS gravity theory can be done considering the $\mathfrak{so}\left(2,2\right)\times\mathfrak{so}\left(2,1\right)$ algebra, also denoted as AdS-Lorentz (AdS-$\mathcal{L}$) algebra \cite{Diaz:2012zza,Fierro:2014lka,Concha:2018jjj}. Applications of the AdS-$\mathcal{L}$ algebra and generalizations can be found in \cite{Concha:2016kdz,Concha:2017nca}, where the higher-dimensional pure Lovelock gravity is recovered as a particular limit of CS gravity theories based on the AdS-$\mathcal{L}$ symmetries.\footnote{The interested reader can also find applications of supersymmetric extensions of the AdS-$\mathcal{L}$ algebra in four spacetime dimensions in the context of supergravity theories formulated à la MacDowell-Mansouri in \cite{Ipinza:2016con,Penafiel:2018vpe,Banaudi:2018zmh}.} Besides, in \cite{Bonanos:2008ez,Bonanos:2008kr,Gomis:2009dm}, deformations of the Maxwell algebra and their dynamics through non-linear realizations were studied.  

At the NR level, the study of a NR version of the three-dimensional CS Maxwell gravity theory was first considered in \cite{Aviles:2018jzw}.  As was discussed in \cite{Aviles:2018jzw}, a finite and non-degenerate NR CS action required to consider the NR limit of a particular $\rm{U}(1)$ enlargement of the Maxwell symmetry leading to a Maxwellian extended Bargmann (MEB) algebra. Interestingly, such symmetry can alternatively be recovered as a vanishing cosmological limit of an enlarged extended Bargmann (EEB) symmetry \cite{Concha:2019lhn} and as a semigroup expansion of the Nappi-Witten algebra \cite{Concha:2019lhn,Penafiel:2019czp}. The generalization of the MEB symmetry and its supersymmetric extension have been subsequently approached in \cite{Concha:2020sjt,Concha:2020ebl} and \cite{Concha:2019mxx,Concha:2020eam}, respectively.

On the other hand, although the Maxwellian UR gravity theory has been partially studied in \cite{Gomis:2019nih}, the corresponding CS action is unknown.
Motivated by the prominent role of Maxwellian algebras in the context of (super)gravity and by the fact that, to our knowledge, the Maxwellian UR gravity model remains unexplored, we present a Maxwellian generalization of the three-dimensional CS Carroll gravity theory. In particular, we would like to study the effect of including, in the UR regime, a covariantly constant electromagnetic field in the three-dimensional CS gravity,
without introducing a cosmological constant.
In this work, we first show that the Maxwell Carroll symmetry can be obtained as an UR limit of the Maxwell algebra. However, similarly to the NR case, an extended version of the Maxwell Carroll is required. Such extension is necessary in order for the invariant tensor to be non-degenerate and thus allow the proper construction of a CS action. Subsequently, we include a cosmological constant to the theory by considering an enlarged Carroll symmetry which can be seen as the corresponding UR counterpart of the AdS-$\mathcal{L}$ symmetry. We show that the model with cosmological constant yielding the Maxwellian extended Carroll CS gravity theory in the limit of vanishing cosmological is based on an enlarged extended version of the Carroll symmetry exhibiting a non-degenerate invariant tensor and therefore allowing the proper construction of a CS gravity theory.

The paper is organized as follows: In Section \ref{Maxgrav} we briefly review the relativistic Maxwell and AdS-$\mathcal{L}$ CS gravity theories in three spacetime dimensions. Sections \ref{MC}, \ref{MEC} and \ref{CS1} contain our main results. In Section \ref{MC} we derive the Maxwellian Carroll algebra by considering the UR limit of the relativistic Maxwell algebra and we write an UR CS gravity action based on the Maxwellian Carroll algebra. In Section \ref{MEC} we consider an extended version of the Maxwellian Carroll symmetry which exhibits a non-degenerate invariant bilinear form and therefore allows the proper construction of a CS Maxwellian extended Carroll gravity theory. In Section \ref{CS1} we show that a cosmological constant can be incorporated to the extended Maxwellian Carroll CS gravity model by considering an enlarged Carroll algebra. We introduce additional bosonic generators to the enlarged Carroll algebra, obtaining an enlarged extended Carroll algebra, in order to have a non-degenerate invariant tensor allowing the proper construction of an UR CS gravity action. In the flat limit, the latter precisely boils down to the Maxwellian extended Carroll CS gravity theory.
Section \ref{conclusions} is devoted to some concluding remarks and possible future developments.


\section{Three-dimensional Maxwell and AdS-Lorentz Chern-Simons gravity}\label{Maxgrav}

In this section, we will briefly review the three-dimensional AdS-$\mathcal{L}$ CS gravity theory \cite{Diaz:2012zza,Fierro:2014lka,Concha:2018jjj}, along with the CS gravity theory obtained after considering its flat limit, known as the Maxwell gravity \cite{Salgado:2014jka,Hoseinzadeh:2014bla,Caroca:2017izc,Concha:2018zeb}.

The AdS-$\mathcal{L}$ algebra can be seen as a semi-simple enlargement of the Poincaré symmetry \cite{Soroka:2006aj}, allowing to incorporate a cosmological constant to the Maxwell gravity. Such enlarged algebra is spanned by the set of generators $\left\{
\tilde{J}_{A},\tilde{P}_{A},\tilde{Z}_{A}\right\} $, which satisfy the non-vanishing commutation relations
\begin{eqnarray}
\left[ \tilde{J}_{A},\tilde{J}_{B}\right] &=&\epsilon _{ABC}\tilde{J}^{C}\,,  \qquad \,
\left[ \tilde{J}_{A},\tilde{P}_{B}\right] =\epsilon _{ABC}\tilde{P}^{C}\,,  \notag \\
\left[ \tilde{J}_{A},\tilde{Z}_{B}\right] &=&\epsilon _{ABC}\tilde{Z}^{C}\,,  \qquad
\left[ \tilde{P}_{A},\tilde{P}_{B}\right] =\epsilon _{ABC}\tilde{Z}^{C}\,, \notag \\
\left[ \tilde{Z}_{A},\tilde{Z}_{B}\right] &=&\frac{1}{\ell ^{2}}\epsilon _{ABC}\tilde{Z}^{C}\,, \quad
\left[ \tilde{P}_{A},\tilde{Z}_{B}\right] =\frac{1}{\ell ^{2}}\epsilon _{ABC}\tilde{P}^{C}\,, \label{Adslor}
\end{eqnarray}%
where $\ell $ is a length parameter related to the (negative) cosmological constant through $\Lambda \propto - \frac{1}{\ell^2}$, the Lorentz indices $A,B,C=0,1,2$ are lowered and raised with
the Minkowski metric $\eta _{AB}=\left( -1,1,1\right) $ and $\epsilon _{ABC}$
is the three-dimensional Levi Civita tensor which satisfies $\epsilon
_{012}=-\epsilon ^{012}=1$. Note that the flat limit $\ell\rightarrow\infty $  applied to the AdS-$\mathcal{L}$ algebra \eqref{Adslor} reproduces the Maxwell symmetry.

The AdS-$\mathcal{L}$ algebra admits the following non-vanishing components of
an invariant tensor of rank two: 
\begin{eqnarray}
\left\langle \tilde{J}_{A}\tilde{J}_{B}\right\rangle &=&\tilde{\alpha}_{0}\eta _{AB}\,, \qquad \, \,
\left\langle \tilde{J}_{A}\tilde{P}_{B}\right\rangle =\tilde{\alpha}_{1}\eta _{AB}\,,  \notag \\
\left\langle \tilde{J}_{A}\tilde{Z}_{B}\right\rangle &=&\tilde{\alpha}_{2}\eta _{AB}\,, \qquad \, \,
\left\langle \tilde{P}_{A}\tilde{P}_{B}\right\rangle =\tilde{\alpha}_{2}\eta _{AB}\,,  \notag \\
\left\langle \tilde{Z}_{A}\tilde{P}_{B}\right\rangle &=&\frac{\tilde{\alpha}_{1}}{\ell ^{2}}\eta
_{AB}\,, \qquad
\left\langle \tilde{Z}_{A}\tilde{Z}_{B}\right\rangle =\frac{\tilde{\alpha}_{2}}{\ell ^{2}}\eta
_{AB}\,,\label{IT}
\end{eqnarray}
where $\tilde{\alpha}_0$, $\tilde{\alpha}_1$ and $\tilde{\alpha}_2$ are arbitrary constants. One can see that the invariant tensor given by \eqref{IT} reduce to the components of the invariant tensor for the Maxwell algebra when $\ell\rightarrow\infty $. Let us notice that both the Maxwellian and AdS-$\mathcal{L}$ invariant tensors are non-degenerate, assuring the vanishing of the curvature two-forms when one considers the field equations of the associated CS gravity theory.

The gauge connection one-form $A$ for the AdS-$\mathcal{L}$ algebra
reads
\begin{equation}
A=\tilde{W}^{A}\tilde{J}_{A}+\tilde{E}^{A}\tilde{P}_{A}+\tilde{K}^{A}\tilde{Z}_{A}\,,  \label{oneform1}
\end{equation}%
where $\tilde{W}^{A}$ denotes the spin-connection, $\tilde{E}^{A}$ is the dreibein and $\tilde{K}^{A} $ is the gauge field associated with the Maxwellian $\tilde{Z}_A$ generator. The
corresponding curvature two-form is given by
\begin{equation}
F=\tilde{R}^{A}\left( \tilde{W}\right) \tilde{J}_{A}+\tilde{R}^{A}\left( \tilde{E}\right) \tilde{P}_{A}+\tilde{R}^{A}\left( \tilde{K}\right)
\tilde{Z}_{A}\,,
\end{equation}
with
\begin{eqnarray}
\tilde{R}^{A}\left( \tilde{W}\right) &:=& d\tilde{W}^{A}+\frac{1}{2}\epsilon ^{ABC}\tilde{W}_{B}\tilde{W}_{C}\,,
\notag \\
\tilde{R}^{A}\left( \tilde{E}\right) &:=& d\tilde{E}^{A}+\epsilon ^{ABC}\tilde{W}_{B}\tilde{E}_{C}+\frac{1}{\ell^2}\epsilon ^{ABC}\tilde{K}_{B}\tilde{E}_{C}\,,  \notag
\\
\tilde{R}^{A}\left( \tilde{K}\right) &:=& d\tilde{K}^{A}+\epsilon ^{ABC}\tilde{W}_{B}\tilde{K}_{C}+\frac{1}{2\ell^2}
\epsilon ^{ABC}\tilde{K}_{B}\tilde{K}_{C}+\frac{1}{2}
\epsilon ^{ABC}\tilde{E}_{B}\tilde{E}_{C}\,.\label{AdSLorCurv}
\end{eqnarray}
Considering the gauge connection one-form (\ref{oneform1}) and the non-vanishing
components of the invariant tensor (\ref{IT}) in the three-dimensional CS
expression,
\begin{equation}
I=\frac{k}{4\pi }\int \left\langle AdA+\frac{2}{3}A^{3}\right\rangle \,,
\label{CS}
\end{equation}
with $k$ being the CS level of the theory related to the gravitational
constant $G$, we find the following CS gravity action for the AdS-$\mathcal{L}$ algebra \cite{Concha:2018jjj}:
\begin{align}
& I_{\text{AdS-}\mathcal{L}}=\frac{k}{4\pi }\int \left[ \tilde{\alpha}_{0}\left(
\tilde{W}^{A}d\tilde{W}_{A}+\frac{1}{3}\,\epsilon ^{ABC}\tilde{W}_{A}\tilde{W}_{B}\tilde{W}_{C}\right) \right.
\notag \\
&  +\left. \tilde{\alpha}_{1}\left( 2\tilde{E}_{A}\tilde{R}^{A}(\tilde{W})+\frac{2}{\ell ^{2}}%
\,\tilde{E}_{A}\tilde{F}^{A}(\tilde{K})+\frac{1}{3\ell ^{2}}\,\epsilon ^{ABC}\tilde{E}_{A}\tilde{E}_{B}\tilde{E}_{C} \right)
\right.  \notag \\
&  +\left. \tilde{\alpha}_{2}\left( \tilde{T}^{A}\tilde{E}_{A}+\frac{1}{ \ell ^{2}}\,\epsilon
^{ABC}\tilde{E}_{A}\tilde{K}_{B}\tilde{E}_{C}+2\tilde{K}_{A}\tilde{R}^{A}(\tilde{W})+\frac{1}{\ell ^{2}}\tilde{K}_{A}\,D_{\tilde{W}}\tilde{K}^{A}+%
\frac{1}{3\ell ^{4}}\epsilon ^{ABC}\tilde{K}_{A}\tilde{K}_{B}\tilde{K}_{C} \right) \right]\,,\label{adsloract}
\end{align}
where we have defined the curvature $\tilde{F}^{A}(\tilde{K}):=D_{\tilde{W}}\tilde{K}^{A}+\frac{1}{2\ell ^{2}%
}\epsilon ^{ABC}\tilde{K}_{B}\tilde{K}_{C}$, being $D_{\tilde{W}}$ the Lorentz covariant
derivative, $D_{\tilde{W}}\Theta ^{A}=d\Theta ^{A}+\epsilon ^{ABC}\tilde{W}_{B}\Theta _{C}.$ Furthermore, we have that $\tilde{T}^{A}:=d\tilde{E}^{A}+\epsilon ^{ABC}\tilde{W}_{B}\tilde{E}_{C}$ is the usual torsion two-form.

From the CS action we can see that it is split in three independent different sectors. The piece along $\tilde{\alpha}_0$ corresponds to the gravitational (or exotic) Lagrangian \cite{Witten:1988hc}, while the Einstein-Hilbert term, a cosmological constant term and a contribution involving the gauge field $\tilde{K}^A$ appear along the $\tilde{\alpha}_1$ constant. Furthermore, the term related to the $\tilde{\alpha}_2$ constant contains a torsional term and also terms involving the additional field $\tilde{K}^A$.
Naturally, the flat limit $\ell\rightarrow\infty$ of the relativistic CS action \eqref{adsloract} leads to the Maxwell CS gravity action \cite{Salgado:2014jka,Hoseinzadeh:2014bla,Caroca:2017izc,Concha:2018zeb}. As it was shown in \cite{Concha:2018zeb, Concha:2018jjj}, the presence of the additional gauge field $\tilde{K}^{A}$ modifies not only the asymptotic sector but also the vacuum of the Maxwell and AdS-$\mathcal{L}$ CS gravity theories, respectively.

Non-relativistic versions of the Maxwell and the AdS-$\mathcal{L}$ algebras have been recently
considered in \cite{Aviles:2018jzw,Concha:2019lhn,Penafiel:2019czp,Concha:2019mxx,Concha:2020sjt,Concha:2020ebl} and have required $\rm{U}(1)$ enlargements in
order to have finite and non-degenerate NR CS gravity actions after the non-relativistic limit. The corresponding  NR algebras are, as their relativistic counterparts, related through the flat limit $\ell\rightarrow\infty$.

In what follows, we will study diverse UR versions of the previously introduced three-dimensional CS gravity theories. First, we will show that a UR limit can be applied to obtain the Carrollian version of the Maxwell and AdS-$\mathcal{L}$ gravity theories. To this aim, we decompose the $A$-index as follows:
\begin{equation}
    A\rightarrow\left(0,a\right), \qquad  a=1,2\,. \label{decom}
\end{equation}
Then, we will apply a particular redefinition to the aforementioned relativistic Maxwell and AdS-$\mathcal{L}$ algebras and we will take the UR limit in order to get the Carrollian version of these algebras. We shall call them as Maxwellian and enlarged Carroll algebras, respectively. We will also consider the contraction at the level of the invariant tensors in order to construct the corresponding UR CS gravity actions. Nevertheless, although the UR symmetries are well-defined in the limit, the respective UR invariant bilinear forms obtained by performing the Carrollian contraction starting from the AdS-$\mathcal{L}$ invariant tensor are degenerate. As we shall see, such drawback can be overcome by considering extensions of the Maxwellian and enlarged Carroll symmetries assuring the proper development of UR CS gravity theories. 


\section{Maxwellian Carroll Chern-Simons gravity theory}\label{MC}

Let us consider now the Carrollian version of the Maxwell algebra, where we recall that the latter can be obtained by taking the $\ell\rightarrow\infty$ limit of \eqref{Adslor}. 

The Maxwellian Carroll algebra is obtained by performing the indices decomposition \eqref{decom} in the Maxwell algebra, and subsequently performing the Carroll contraction, in view of which we introduce the $\sigma$ parameter by the redefinition 
\begin{equation}\label{redef}
    \tilde{J}_{a} \rightarrow \sigma K_{a} \,, \quad \tilde{P}_{0}\rightarrow\sigma H \,, \quad \tilde{Z}_{a}\rightarrow\sigma Z_{a} \,, \quad 
    \tilde{J}_{0}\rightarrow J\,, \quad
    \tilde{P}_{a}\rightarrow P_{a}\,, \quad \tilde{Z}_{0}\rightarrow Z\,.
\end{equation}
Then, considering the limit $\sigma\rightarrow\infty$,\footnote{The $\sigma$ parameter is related to the speed of light $c$ as $\sigma\rightarrow 1/c$ such that the limit $\sigma\rightarrow\infty$ corresponds to the ultra-relativistic limit, $c \rightarrow 0$.}. we obtain the algebra generated by the set of generators $\lbrace J, K_{a}, H,P_{a},Z,Z_{a}\rbrace$,
satisfying the non-vanishing commutation relations
\begin{eqnarray}
\left[ J,K_{a}\right] &=&\epsilon _{ab}{K}_{b}\,, \qquad \qquad 
\left[ J,P_{a}\right] =\epsilon _{ab}{P}_{b}\,, \qquad \qquad
\left[K_{a},{P}_{b}\right] =-\epsilon _{ab}H\,,  \notag
\\
\left[ {J},{Z}_{a}\right] &=&\epsilon _{ab}{Z}_{b}\,,\qquad \quad \ \
\left[ P_{a},P_{b}\right] =-\epsilon _{ab}Z\,,\qquad \quad \ \,
\left[ H,P_{a}\right] =\epsilon _{ab}Z_{b}\,,  \notag
\\
\left[ K_{a},Z\right] &=&-\epsilon _{ab}Z_{b}\,,  \label{MCA}
\end{eqnarray}
where $a=1,2$, $\epsilon_{ab}\equiv\epsilon_{0ab}$,  $\epsilon^{ab}\equiv\epsilon^{0ab}$. Here $\lbrace J, K_{a}, H,P_{a},Z,Z_{a}\rbrace$ are spacial rotations, Carrollian boosts, time translations, space translations and the UR Maxwellian generators $Z$ and $Z_{a}$, respectively. Note that the three commutators appearing in the first line of \eqref{MCA} define the Carroll algebra \cite{LevyLeblond, Bacry:1968zf}, which is obtained by a Carroll contraction of the Poincaré algebra and appears by suppressing the Maxwell generators $Z$ and $Z_a$. Analogously to the Maxwellian extended Bargmann symmetry \cite{Aviles:2018jzw}, one can notice that the $Z$ generator present in this Maxwellian generalization of the Carroll algebra is not a central charge and thus does not appear as a central extension of the Carroll algebra. On the other hand, the Maxwellian Carroll algebra presented here is contained as a particular case of the infinite-dimensional Carrollian Maxwell algebra introduced in \cite{Gomis:2019nih}, which was obtained applying the $S$-expansion procedure \cite{Izaurieta:2006zz}.

The corresponding gauge connection one-form reads
\begin{equation}
    A=\tau H+e^{a}P_{a}+\omega J+\omega^{a}K_{a}+kZ+k^{a}Z_{a}\,. \label{oneformMCA}
\end{equation}
The curvature two-form $F=dA+\frac{1}{2}\left[ A,A\right] $ is given by%
\begin{eqnarray}
F =R\left(\tau\right) H+R^{a}\left(e^{b}\right)P_{a}+R\left(\omega\right) J+R^{a}\left(\omega^{b}\right)K_{a}+R\left(k\right)Z+R^{a}\left(k^{b}\right)Z_{a}\,, \label{curvMC}
\end{eqnarray}
where the UR curvatures assiociated with the Maxwellian Carroll algebra are defined as follows:
\begin{eqnarray}
R\left( \tau \right) &=&d\tau+\epsilon^{ab}\omega_{a}e_{b} \,, \qquad \qquad R^{a}\left( e ^{b}\right)  = de ^{a}+\epsilon ^{ac}\omega
e_{c}\,,  \notag \\
R\left( \omega \right) &=& d\omega \,, \qquad \qquad \qquad \qquad R^{a}\left( \omega^{b}\right) = d \omega^a + \epsilon ^{ac}\omega \omega_{c} \,,  \notag \\
R\left( k\right) &=& dk+\frac{1}{2}\epsilon^{ab}e_{a}e_{b} \,, \qquad \quad \,\, R^{a}\left( k^{b}\right) = d k^a + \epsilon ^{ac}\omega k_{c} + \epsilon
^{ac}\tau e_{c} + \epsilon ^{ac} k \omega_{c} \,.  \label{curvMCA}
\end{eqnarray}
Now, in order to construct the CS action invariant under the Maxwellian Carroll algebra, we also require the non-vanishing components of the invariant bilinear form. These can be derived as the Carroll contraction of the non-vanishing components of the invariant tensor for the Maxwell algebra, where the latter can be obtained from \eqref{IT} considering $\ell\rightarrow\infty$. Note that, in order to end up with a non-trivial invariant tensor for the contracted algebra, we also have to rescale the relativistic constants appearing in \eqref{IT} as follows:
\begin{equation}
    \tilde{\alpha}_{0}\rightarrow\alpha_{0}\,,\qquad \tilde{\alpha}_{1}\rightarrow\sigma\alpha_{1}\,,\qquad \tilde{\alpha}_{2}\rightarrow\alpha_{2}\,. \label{rescale}
\end{equation}
Then, after taking the limit $\sigma\rightarrow\infty$, we obtain the following non-vanishing components of the invariant bilinear form for the Maxwellian Carroll algebra:
\begin{eqnarray}
\left\langle JJ\right\rangle &=&-\alpha _{0}\,, \qquad \qquad \qquad \left\langle J H\right\rangle =-\alpha _{1}\,, \qquad \qquad \qquad \left\langle J Z\right\rangle =-\alpha _{2}\,, \notag \\
\left\langle K_{a}P_{b}\right\rangle &=& \alpha _{1}\delta_{ab} \,,  \qquad \qquad \quad
\left\langle P_{a}P_{b}\right\rangle = \alpha _{2}\delta_{ab} \,.\label{ITMCA}
\end{eqnarray}
Then, considering the gauge connection one-form for the Maxwellian Carroll algebra \eqref{oneformMCA} and the non-vanishing components \eqref{ITMCA} in the general expression of the CS action (\ref{CS}), we find the following UR CS action for the Maxwellian Carroll  algebra:
\begin{equation}\label{MCAact}
I_{{\text{Maxwell Carroll}}}=\frac{k}{4\pi}\int \bigg \lbrace -\alpha_{0}\omega R\left(\omega\right)+\alpha_{1}\left[-2\tau R\left(\omega\right)+2e_{a}R^{a}\left(\omega^{b}\right)\right]  +\alpha_{2}\left[-2k R\left(\omega\right)+e_{a}R^{a}\left(e^{b}\right)\right] \bigg \rbrace \,. 
\end{equation}
This novel UR CS gravity action is defined by three different independent sectors, each one being invariant under the Maxwellian Carroll algebra. The first term along the $\alpha_{0}$ constant corresponds to the UR version of the exotic Lagrangian, while the term proportional to $\alpha_{1}$ is the usual Carrollian gravity \cite{Bergshoeff:2016soe}. On the other hand, the term along $\alpha_{2}$ corresponds to the novel Maxwellian Carroll contribution. Let us notice that the UR CS gravity action \eqref{MCAact} can also be recovered from the relativistic Maxwell CS action, the latter being the flat limit of \eqref{adsloract}, by performing the following redefinition of the one-form gauge fields:
\begin{equation}\label{redef2}
   \tilde{W}^{a} \rightarrow \frac{1}{\sigma} \omega^{a}\,, \quad  \tilde{E}^{0}\rightarrow\frac{1}{\sigma} \tau \,, \quad \tilde{K}^{a}\rightarrow\frac{1}{\sigma} k^{a}\,, \quad \tilde{W}^{0}\rightarrow \omega\,, \quad \tilde{E}^{a}\rightarrow e^{a}\,, \quad \tilde{K}^{0}\rightarrow k\,,
\end{equation}
together with \eqref{rescale}, and consequently taking the limit $\sigma \rightarrow \infty$ directly at the level of the action.

Although the Carroll contraction of the Maxwell algebra allows the construction of a finite UR CS gravity action, it does not avoid the degeneracy problem. Indeed, the Maxwellian Carroll algebra \eqref{MCA} only allows for a degenerate invariant bilinear form. In particular, in our construction the absence of the $k_a$ gauge field, and therefore of its CS kinetic term, in the CS action \eqref{MCAact} is due to the degeneracy of the invariant tensor. In the present case the Carrollian limit procedure led to an algebra with a degenerate bilinear form that does not reproduce a well-defined CS action, as \eqref{MCAact} does not contain a kinetic term for each gauge field. Then, as was mentioned in \cite{Aviles:2018jzw}, guaranteeing finiteness of the action in the contraction process does not guarantee a non-degenerate bilinear form. Then, a natural question is if there exist another Carrollian version of the Maxwell algebra allowing for a non-degenerate invariant tensor. The answer to this question will be given in the next section.

\section{Maxwellian Extended Carroll Chern-Simons gravity theory}\label{MEC}

A different Carrollian version of the Maxwell algebra is contained as a particular case in the infinite-dimensional Carrollian Maxwell algebra introduced in \cite{Gomis:2019nih}. The algebra is generated by the set of generators $\lbrace J, K_{a}, H,P_{a},Z,Z_{a},S,L_{a},T\rbrace$,
satisfying the non-vanishing commutation relations
\begin{eqnarray}
\left[ J,K_{a}\right] &=&\epsilon _{ab}{K}_{b}\,, \qquad \qquad 
\left[ J,P_{a}\right] =\epsilon _{ab}{P}_{b}\,, \qquad \quad \ 
\left[K_{a},{P}_{b}\right] =-\epsilon _{ab}H\,,  \notag
\\
\left[ {J},{Z}_{a}\right] &=&\epsilon _{ab}{Z}_{b}\,,\qquad \quad \ \
\left[ P_{a},P_{b}\right] =-\epsilon _{ab}Z\,,\qquad \quad \,
\left[ H,P_{a}\right] =\epsilon _{ab}Z_{b}\,,  \notag
\\
\left[ K_{a},Z\right] &=&-\epsilon _{ab}Z_{b}\,,\qquad \ \, \left[ K_{a},K_{b}\right] =-\epsilon _{ab}S\,,\qquad \quad \ \, \left[ J,L_{a}\right] =\epsilon _{ab}L_{b}\,,
\notag
\\
\left[ {S},{P}_{a}\right] &=&\epsilon _{ab}{L}_{b}\,, \qquad \quad \ \, \left[ K_{a},Z_{b}\right] =-\epsilon _{ab}T\,, \qquad \quad  \left[ P_{a},L_{b}\right] =-\epsilon _{ab}T\,,  \notag
\\
\left[ K_{a},H\right] &=&-\epsilon _{ab}L_{b}\,.\label{MECA}
\end{eqnarray}
We shall call the algebra \eqref{MECA} as Maxwellian Extended Carroll (MEC) algebra.\footnote{The MEC algebra can be obtained from the relativistic Maxwell algebra by performing a particular expansion (see \cite{Gomis:2019nih}, where the MEC algebra corresponds to a finite sub-case of the infinite-dimensional Maxwellian UR algebra constructed).}. The MEC algebra is characterized by the presence of the $S$ and $L_a$ generators along with a central charge $T$.  Note that it reduces to the Maxwellian Carroll algebra when $S=T=L_{a}=0$.
The algebra \eqref{MECA} admits the following non-vanishing components of a non-degenerate invariant tensor:
\begin{equation}\label{ITMECA}
    \begin{split}
        & \left\langle JJ\right\rangle = \left\langle JS\right\rangle=-\alpha _{0}\,, \qquad \qquad \qquad \qquad \quad \, \left\langle K_{a} K_{b}\right\rangle = \alpha _{0}\delta_{ab} \,, \\
        & \left\langle J H\right\rangle = -\alpha_{1}\,, \quad \qquad \qquad \qquad \qquad \qquad \quad \,\,\, \left\langle K_{a}P_{b}\right\rangle = \alpha _{1}\delta_{ab}\,, \\
        & \left\langle J Z\right\rangle =\left\langle J T\right\rangle=\left\langle S Z\right\rangle=\left\langle H H\right\rangle=-\alpha _{2}\,, \quad \quad \left\langle P_{a}P_{b}\right\rangle = \left\langle P_{a}L_{b}\right\rangle=\left\langle K_{a}Z_{b}\right\rangle = \alpha _{2}\delta_{ab}\,,
    \end{split}
\end{equation}
where $\alpha_0$, $\alpha_1$ and $\alpha_2$ are arbitrary constants. Let us note that here non-degeneracy requires a Maxwellian extension of the Carroll algebra analogously to the non-relativistic case in which a Maxwellian extension of the Bargmann (MEB) algebra assures the non-degeneracy of the invariant tensor \cite{Aviles:2018jzw}. Nevertheless, unlike Carroll and Bargmann algebra, there is no duality between the MEC and the MEB algebras since they differ in the amount of generators. Although one could see some similarity with the generalized Maxwellian extended Bargmann (GMEB) algebra introduced in \cite{Concha:2020sjt}, they are also quite different at the level of amount of generators. The following table summarizes the generators content of the previously mentioned UR and non-relativistic versions of the Maxwell algebra:
\begin{equation}
    \begin{tabular}{|m{6,5em}||c|c|c|c|}
\hline
 & Maxwell Carroll & MEC & MEB & GMEB \\ \hline\hline
Time generators & $J,\ H,\ Z$ & $J,\ H,\ Z,\ S$ & $J,\ H,\ Z,$  & $J,\ H,\ Z,\ N$ \\ \hline
Spatial generators & $K_a,\ P_a,\ Z_a$ & $K_a,\ P_a,\ Z_a,\ L_a$ & $G_a,\ P_a,\ Z_a$ & $G_a,\ P_a,\ Z_a,\ N_a$ \\ \hline
Central charges &  & $T$ & $S,\ M,\ T$ & $S,\ M,\ T,\ V$ \\ \hline
Amount of generators & $9$ & $13$ & $12$ & $16$ \\ \hline
\end{tabular}\notag
\end{equation}
\begin{center}
   Table 1: Generators of UR and NR versions of the Maxwell algebra.
\end{center}

To construct an UR CS action gauge-invariant under the MEC algebra \eqref{MECA}, let us consider the following gauge connection one-form:
\begin{equation}
    A=\tau H+e^{a}P_{a}+\omega J+\omega^{a}K_{a}+kZ+k^{a}Z_{a}+sS+tT+l^{a}L_{a}\,. \label{oneformMECA}
\end{equation}
The corresponding curvature two-form $F=dA+\frac{1}{2}\left[ A,A\right] $ is given by
\begin{eqnarray}
F&=&R\left(\tau\right) H+R^{a}\left(e^{b}\right)P_{a}+R\left(\omega\right) J+R^{a}\left(\omega^{b}\right)K_{a}+R\left(k\right)Z+R^{a}\left(k^{b}\right)Z_{a} \notag \\
&&+R\left(s\right)S+R\left(t\right)T+R^{a}\left(l^{b}\right)L_{a}\,, \label{curvMEC}
\end{eqnarray}
where the respective field-strengths of the gauge fields dual to the generators of the MEC algebra are given by \eqref{curvMCA} along with
\begin{eqnarray}
R\left( s\right) &=&ds+\frac{1}{2}\epsilon^{ab}\omega_{a}\omega_{b}\,, \notag \\
R\left( t\right) &=&dt+\epsilon^{ab}\omega_{a}k_{b}+\epsilon^{ab}e_{a}l_{b}\,, \notag \\
R^{a}\left(l^{b}\right)&=&dl^{a}+\epsilon^{ab}\omega l_{b}+\epsilon^{ab}s e_{b}+\epsilon^{ab}\tau \omega_{b}\,.  \label{curvMECA}
\end{eqnarray}

Considering the gauge connection one-form for the MEC algebra \eqref{oneformMECA} and the non-vanishing components of the invariant tensor \eqref{ITMECA} in the general
expression of the CS action (\ref{CS}), we find the following UR CS action for the MEC algebra,
\begin{eqnarray}
I_{\text{MEC}}&=&\frac{k}{4\pi}\int \bigg \lbrace \alpha_{0}\left[-\omega R\left(\omega\right)-2s R\left(\omega\right)+\omega_{a}R^a \left(\omega^{b}\right)\right]+\alpha_{1}\left[-2\tau R\left(\omega\right)+2e_{a}R^{a}\left(\omega^{b}\right)\right]   \notag \\
&&+\alpha_{2}\left[-2k R\left(\omega\right)+e_{a}R^{a}\left(e^{b}\right)-2s R\left(k\right)-\tau R\left(\tau\right)-2t R\left(\omega\right)\right. \notag \\
&&+ \left.2l_{a}R^{a}\left(e^{b}\right)+\omega_{a}R^{a}\left(k^{b}\right)+k_{a}R^{a}\left(\omega^{b}\right)\right] \bigg \rbrace \,.\label{MECAact}
\end{eqnarray}

This novel UR CS gravity action is defined by three different independent sectors, each one being invariant under the MEC algebra. Note that, as in the previous case, the term proportional to $\alpha_{1}$ is the usual Carrollian gravity \cite{Bergshoeff:2016soe}. However, unlike the Maxwell Carroll CS gravity, the $k^{a}$ gauge field appears explicitly along the $\alpha_2$ constant. The term along $\alpha_{2}$ describes a new UR Maxwell gravity which requires the introduction of an additional gauge field $l^{a}$ and two one-form gauge fields $s$ and $t$, respectively dual to the generators $L_a$, $S$ and $T$ of the UR algebra \eqref{MECA}. Note that the MEC algebra allows for a non-degenerate invariant bilinear form which implies that in the action there is a kinetic term of each gauge field and that the field equations of the theory correspond to the vanishing of the MEC field-strengths. Indeed, the equations of motion are given by 
\begin{eqnarray}
 \delta \omega_a&:& \qquad \alpha_0 R^a\left(\omega^b\right) + \alpha_1 R^a\left(e^b\right) + \alpha_2  R^a\left(k^b\right) =0\,, \notag \\
 \delta \omega&:& \qquad \alpha_0 \left[R\left(\omega\right)+R\left(s\right)\right] + \alpha_1 R\left(\tau\right) + \alpha_2\left[ R\left(k\right)+R\left(t\right)\right] =0\,, \notag \\
 \delta e_a&:& \qquad \alpha_1 R^a\left(\omega^b\right) + \alpha_2 \left[ R^a\left(e^b\right) +  R^a\left(l^b\right)\right] =0\,, \notag \\
 \delta \tau&:& \qquad \alpha_1 R\left(\omega\right)+\alpha_2 R\left(\tau\right) =0\,, \notag \\
 \delta k_a&:& \qquad \alpha_2 R^a\left(\omega^b\right)=0 \,, \notag \\
 \delta k &:& \qquad \alpha_2 \left[R\left(\omega\right)+R\left(s\right)\right]=0 \,, \notag \\
 \delta s &:& \qquad \alpha_0 R\left(\omega\right) + \alpha_2  R\left(k\right) =0\,, \notag \\
 \delta t &:& \qquad \alpha_2 R\left(\omega\right)=0\,, \notag \\
 \delta l_a &:& \qquad \alpha_2R^a\left(e^b\right)=0\,.   \label{eomMECA}
\end{eqnarray}
As the non-degeneracy of the invariant tensor \eqref{ITMECA} requires $\alpha_2\neq0$, the field equations \eqref{eomMECA} are equivalent to the vanishing of the curvature two-forms \eqref{curvMCA} along with \eqref{curvMECA}.

\section{Enlarged Carroll Chern-Simons gravity theory and extension}\label{CS1}

Analogously to the non-relativistic Maxwell gravity, the MEC CS gravity theory does not contain a cosmological constant. The inclusion of a cosmological constant in the UR CS gravity theory can be done considering the AdS-Carroll symmetry \cite{Matulich:2019cdo}. Here, we show that a cosmological constant can be incorporated to the MEC CS gravity model by considering an enlarged Carroll algebra. Nevertheless, as in the Maxwell case, an extended UR symmetry is required in order to avoid degeneracy. For completeness, we first explore the enlarged Carroll CS gravity along with its flat limit.
 
\subsection{Enlarged Carroll Chern-Simons gravity}\label{EnlCar}

An enlarged Carroll algebra can be derived applying the Carroll contraction to the AdS-Lorentz algebra \eqref{Adslor}. Indeed, performing the redefinition of the generators as in \eqref{redef} and after taking the limit $\sigma\rightarrow\infty$, we get 
\begin{eqnarray}
\left[ J,K_{a}\right] &=&\epsilon _{ab}{K}_{b}\,, \qquad \qquad 
\left[ J,P_{a}\right] =\epsilon _{ab}{P}_{b}\,, \qquad \qquad
\left[K_a,{P}_{b}\right] =-\epsilon _{ab}H\,,  \notag
\\
\left[ {J},{Z}_{a}\right] &=&\epsilon _{ab}{Z}_{b}\,,\qquad \quad \ \ 
\left[ P_{a},P_{b}\right] =-\epsilon _{ab}Z\,,\qquad \quad \ \ \,
\left[ H,P_{a}\right] =\epsilon _{ab}Z_{b}\,,  \notag
\\
\left[ K_{a},Z\right] &=&-\epsilon _{ab}Z_{b}\,, \qquad \quad  \, \left[ P_{a},Z\right] =-\frac{1}{\ell^2}\epsilon _{ab}P_{b}\,, \qquad \  \left[ P_{a},Z_{b}\right] =-\frac{1}{\ell^2}\epsilon _{ab}H\,,  \notag
\\
\left[ Z,Z_{a}\right]& =&\frac{1}{\ell^2}\epsilon _{ab}Z_{b}\,,\label{ECA}
\end{eqnarray}
where $\ell$ is a length parameter related to the inverse of the cosmological constant $\Lambda$. The present symmetry corresponds to the three-dimensional version of the UR symmetry denoted as $\mathcal{C}\mathfrak{L}_4$ in \cite{RUBIO:2020mnh}. Let us note that, as its relativistic counterpart, the Maxwell Carroll algebra \eqref{MCA} appears in the vanishing cosmological constant limit $\ell\rightarrow\infty$.

The gauge connection one-form for the enlarged Carroll symmetry reads
\begin{equation}
    A=\tau H+e^{a}P_{a}+\omega J+\omega^{a}K_{a}+kZ+k^{a}Z_{a}\,. \label{oneformEC}
\end{equation}
The corresponding curvature two-form is given by
\begin{eqnarray}
F =\hat{R}\left(\tau\right) H+\hat{R}^{a}\left(e^{b}\right)P_{a}+\hat{R}\left(\omega\right) J+\hat{R}^{a}\left(\omega^{b}\right)K_{a}+\hat{R}\left(k\right)Z+\hat{R}^{a}\left(k^{b}\right)Z_{a}\,, \label{curvEC}
\end{eqnarray}
where
\begin{eqnarray}
\hat{R}\left( \tau \right) &=&d\tau+\epsilon^{ab}\omega_{a}e_{b}+\frac{1}{\ell^2}\epsilon^{ab}e_{a}k_b \,, \qquad \hat{R}^{a}\left( e ^{b}\right) =de ^{a}+\epsilon ^{ac}\omega
e_{c}+\frac{1}{\ell^2}\epsilon^{ac}ke_{c}\,,  \notag \\
\hat{R}\left( \omega \right) &=& d\omega \,,  \qquad \qquad \qquad \qquad \qquad \quad \,\, \hat{R}^{a}\left( \omega^{b}\right) = d \omega^a + \epsilon ^{ac}\omega \omega_{c} \,,  \notag \\
\hat{R}\left( k\right) &=& dk+\frac{1}{2}\epsilon^{ab}e_{a}e_{b} \,, \notag \\
\hat{R}^{a}\left( k^{b}\right) &=& d k^a + \epsilon ^{ac}\omega k_{c} + \epsilon
^{ac}\tau e_{c} + \epsilon ^{ac} k \omega_{c}+\frac{1}{\ell^2}\epsilon^{ac}kk_{c} \,.  \label{curvECA}
\end{eqnarray}
Let us note that the enlarged Carroll curvatures coincide with the Maxwell ones \eqref{curvMCA} in the flat limit $\ell\rightarrow\infty$. On the other hand, one can show that the enlarged Carroll algebra \eqref{ECA} admits the following non-vanishing components of an invariant tensor:
\begin{eqnarray}
\left\langle JJ\right\rangle &=&-\alpha _{0}\,, \qquad \qquad \qquad \quad \,
\left\langle J H\right\rangle =-\alpha _{1}\,,  \notag \\
\left\langle K_{a}P_{b}\right\rangle &=& \alpha _{1}\delta_{ab}\,,  \qquad \qquad \qquad \ \ \  
\left\langle J Z\right\rangle =-\alpha _{2}\,, \notag \\
\left\langle P_{a}P_{b}\right\rangle &=& \alpha _{2}\delta_{ab} \,, \qquad \qquad \qquad \  \
\left\langle Z H \right\rangle =-\frac{\alpha_1}{\ell^2}\,, \notag \\
\left\langle Z_{a}P_{b}\right\rangle &=& \frac{\alpha _{1}}{\ell^2}\delta_{ab}\qquad \qquad \qquad \ \ \ 
\left\langle Z Z\right\rangle =-\frac{\alpha_2}{\ell^2}\,,  \label{ITECA}
\end{eqnarray}
where the constants $\alpha_0$, $\alpha_1$ and $\alpha_2$ have been obtained by applying the rescaling \eqref{rescale} and the UR limit to the ones appearing in the relativistic invariant tensor \eqref{IT}. Then, considering the gauge connection one-form for the enlarged Carroll algebra \eqref{ECA} and the non-vanishing components of the invariant
tensor \eqref{ITECA} in the general
expression of the CS action (\ref{CS}), we find the following UR CS action:
\begin{eqnarray}
I_{\text{Enlarged Carroll}}&=&\frac{k}{4\pi}\int \bigg \lbrace -\alpha_{0}\omega \hat{R}\left(\omega\right)+2\alpha_{1}\left[-\tau \hat{R}\left(\omega\right)+e_{a}\hat{R}^{a}\left(\omega^{b}\right)+\frac{1}{\ell^2}e_{a}F^{a}\left(k^{b}\right)-\frac{1}{\ell^2}\tau \hat{R}\left(k\right)\right]   \notag \\
&&+\alpha_{2}\left[-2k \hat{R}\left(\omega\right)+e_{a}\hat{R}^{a}\left(e^{b}\right)-\frac{1}{\ell^2}kdk\right] \bigg \rbrace \,, \label{ECAact}
\end{eqnarray}
where $F^{a}\left(k^{b}\right)\equiv dk^{a}+\epsilon^{ac}\omega k_{c}+\epsilon^{ac} k \omega_{c}+\frac{1}{\ell^2}\epsilon^{ac}k k_{c}$. The CS action \eqref{ECAact} is invariant under the enlarged Carroll algebra \eqref{ECA} and reproduces the Maxwellian Carroll CS action \eqref{MCAact} in the vanishing cosmological constant limit $\ell\rightarrow\infty$. Such CS action can alternatively be obtained from the relativistic AdS-Lorentz CS action \eqref{adsloract} by considering the redefinition \eqref{redef2}, together with the rescaling \eqref{rescale}, and taking the limit $\sigma\rightarrow\infty$. Although the Carroll limit is well-defined and the CS action is finite, the enlarged Carroll symmetry only admits a degenerate invariant bilinear form. As in the Maxwellian case, the degeneracy problem can be overcome by considering an extended version of the enlarged Carroll algebra.


\subsection{Enlarged Extended Carroll Chern-Simons gravity action and flat limit}\label{CS2}

The inclusion of a cosmological constant to the non-degenerate UR version of the Maxwell CS gravity theory can be done introducing additional bosonic generators to the enlarged Carroll algebra \eqref{ECA}. Let us consider the set of generators $\{J,K_a,H,P_a,Z,Z_a,S,L_a,T\}$ which satisfy the commutation relations of the MEC algebra \eqref{MECA} along with
\begin{eqnarray}
\left[ H,Z_{a}\right] &=&\frac{1}{\ell^2}\epsilon _{ab}{L}_{b}\,, \qquad \qquad 
\left[ Z,P_{a}\right] =\frac{1}{\ell^2}\epsilon _{ab}{P}_{b}\,, \qquad \quad \ \,
\left[P_{a},{Z}_{b}\right] =-\frac{1}{\ell^2}\epsilon _{ab}H\,,  \notag
\\
\left[ Z,{Z}_{a}\right] &=&\frac{1}{\ell^2}\epsilon _{ab}{Z}_{b}\,,\qquad \qquad
\left[ T,P_{a}\right] =\frac{1}{\ell^2}\epsilon _{ab}{L}_{b}\,, \qquad \quad \ 
\left[ Z_{a},Z_{b}\right] =-\frac{1}{\ell^2}\epsilon _{ab}T\,,\notag \\
\left[ Z, L_{a}\right] &=&\frac{1}{\ell^2}\epsilon_{ab}L_{b}\,,\label{EECA}
\end{eqnarray}
where $\ell$ is related to the inverse of the cosmological constant $\Lambda$. We denote the UR algebra \eqref{EECA} as the Enlarged Extended Carroll (EEC) algebra. Naturally, the EEC algebra reproduces the enlarged Carroll one by setting $S=T=L_a=0$. Moreover, the MEC algebra is recovered in the vanishing cosmological constant limit $\ell\rightarrow\infty$. Observe also that, in \eqref{EECA}, $T$ is not a central charge, while it reduces to a central generator in the $\ell \rightarrow \infty$ limit. One can notice that the EEC algebra admits a non-degenerate invariant tensor whose non-vanishing components are given by \eqref{ITMECA} along with 
\begin{eqnarray}
\left\langle ZH\right\rangle &=&-\frac{\alpha _{1}}{\ell^2}\,, \qquad \qquad \qquad \left\langle Z_{a} P_{b}\right\rangle = \frac{\alpha _{1}}{\ell^2}\delta_{ab} \,,  \notag \\
\left\langle Z Z\right\rangle &=&\left\langle Z T \right\rangle=-\frac{\alpha_{2}}{\ell^2}\,, \qquad \,\, \left\langle Z_{a}Z_{b}\right\rangle = \frac{\alpha _{2}}{\ell^2}\delta_{ab}\,, \label{ITEECA}
\end{eqnarray}
where the non-degeneracy requires $\alpha_2\neq0$ and $\alpha_0\neq\alpha_2\ell^2.$ On the other hand, the EEC gauge connection one-form reads
\begin{equation}
    A=\tau H+e^{a}P_{a}+\omega J+\omega^{a}K_{a}+kZ+k^{a}Z_{a}+sS+tT+l^{a}L_{a}\,. \label{oneformEECA}
\end{equation}
The curvature two-form is given by
\begin{eqnarray}
F&=&\hat{R}\left(\tau\right) H+\hat{R}^{a}\left(e^{b}\right)P_{a}+\hat{R}\left(\omega\right) J+\hat{R}^{a}\left(\omega^{b}\right)K_{a}+\hat{R}\left(k\right)Z+\hat{R}^{a}\left(k^{b}\right)Z_{a} \notag \\
&&+\hat{R}\left(s\right)S+\hat{R}\left(t\right)T+\hat{R}^{a}\left(l^{b}\right)L_{a}\,, \label{curvEEC}
\end{eqnarray}
where the respective field-strengths are given by \eqref{curvECA} and 
\begin{eqnarray}
\hat{R}\left( s\right) &=&ds+\frac{1}{2}\epsilon^{ab}\omega_{a}\omega_{b}\,, \qquad \quad \hat{R}\left( t\right) =dt+\epsilon^{ab}\omega_{a}k_{b}+\epsilon^{ab}e_{a}l_{b}+\frac{1}{2\ell^2}\epsilon^{ab}k_{a}k_{b}\,, \notag \\
\hat{R}^{a}\left(l^{b}\right)&=&dl^{a}+\epsilon^{ab}\omega l_{b}+\epsilon^{ab}s e_{b}+\epsilon^{ab}\tau \omega_{b}+\frac{1}{\ell^2}\epsilon^{ab}\tau k_{b}+\frac{1}{\ell^2}\epsilon^{ab}te_{b}+\frac{1}{\ell^2}\epsilon^{ab}kl_{b}\,.  \label{curvEECA}
\end{eqnarray}
Naturally, the MEC curvatures \eqref{curvMECA} are recovered in the flat limit $\ell\rightarrow\infty$. The UR CS action based on the EEC algebra is obtained by taking into account the non-vanishing components of the invariant tensor \eqref{ITEECA} and the gauge connection one-form \eqref{oneformEECA} into the general CS expression \eqref{CS}. By doing so, we find the following UR CS gravity action invariant under the EEC algebra:
\begin{eqnarray}
I_{\text{EEC}}&=&\frac{k}{4\pi}\int \bigg \lbrace \alpha_{0}\left[-\omega \hat{R}\left(\omega\right)-2s \hat{R}\left(\omega\right)+\omega_{a}\hat{R}^a \left(\omega^{b}\right)\right] \notag \\
&&+2\alpha_{1}\left[-\tau \hat{R}\left(\omega\right)+e_{a}\hat{R}^{a}\left(\omega^{b}\right)+\frac{1}{\ell^2}e_{a}F^{a}\left(k^{b}\right)-\frac{1}{\ell^2}\tau \hat{R}\left(k\right)\right]   \notag \\
&&+\alpha_{2}\left[-2k \hat{R}\left(\omega\right)+e_{a}\hat{R}^{a}\left(e^{b}\right)-2s \hat{R}\left(k\right)-\tau \hat{R}\left(\tau\right)-2t \hat{R}\left(\omega\right)+2l_{a}\hat{R}^{a}\left(e^{b}\right)\right. \notag \\
&&+ \left.\omega_{a}\hat{R}^{a}\left(k^{b}\right)+k_{a}\hat{R}^{a}\left(\omega^{b}\right)-\frac{1}{\ell^2}kdk+\frac{1}{\ell^2}k_a\hat{R}^{a}\left(k^{b}\right)-\frac{2}{\ell^2}t\hat{R}\left(k\right)\right] \bigg \rbrace \,,\label{MEEAact}
\end{eqnarray}
where $F^{a}\left(k^{b}\right)\equiv dk^{a}+\epsilon^{ac}\omega k_{c}+\epsilon^{ac} k \omega_{c}+\frac{1}{\ell^2}\epsilon^{ac}k k_{c}$. As in the enlarged Carroll CS action \eqref{ECAact}, the UR Maxwellian gauge field $k$ and $k_a$ appear in both the $\alpha_1$ and $\alpha_2$ sectors. Nevertheless, it is the very presence of the additional gauge fields which allows to avoid the degeneracy problem present in the enlarged Carroll case. Indeed, for $\alpha_2\neq0$ and $\alpha_0\neq\alpha_2\ell^2$, the field equations exactly correspond to the vanishing of the curvature two-forms \eqref{curvEEC}. Furthermore, in the vanishing cosmological constant limit, the CS action reduces to the non-degenerate MEC gravity action \eqref{MECAact}. 

It is important to mention that, unlike the enlarged Carroll symmetry, the EEC algebra has not been obtained as a Carrollian limit of a relativistic symmetry. It would be worth it to explore if, analogously to the MEC algebra \cite{Gomis:2019nih}, the present UR algebra and its non-degenerate invariant tensor can be alternatively recovered as an UR expansion of a relativistic algebra and its invariant tensor, respectively.


\section{Conclusions}\label{conclusions}

In this work we have presented novel UR symmetries being Maxwellian generalizations and enlargements of the Carroll symmetry. In the Maxwellian case, we have shown that an extended version of the Maxwell Carroll algebra, which we have denoted as MEC algebra, is required in order to avoid degeneracy. Such extension of the UR Maxwell algebra involves the presence of extra bosonic generators ($S,L_a,T$, where in particular $T$ is a central charge), with respect to the Maxwellian Carroll algebra, necessary in order to produce a non-degenerate UR invariant tensor. The CS actions for the Maxwell Carroll and the MEC algebras have been constructed. In particular, the non-degeneracy of the invariant trace in the MEC case ensures that the field equations arising from the MEC gravity theory are given by the vanishing of the curvature two-forms. Consequently, we have studied the origin of the Maxwell Carroll and the MEC gravity theories as a flat limit, showing that both can be recovered in the vanishing cosmological constant limit (namely for $\ell \rightarrow \infty$) of an enlarged Carroll and enlarged extended Carroll (EEC) gravity theories, respectively. Although the enlarged Carroll algebra has a well-defined vanishing cosmological constant limit, it does not admit a non-degenerate invariant tensor. Remarkably, the EEC algebra admits a non-degenerate invariant bilinear form and therefore allows the proper construction of a CS gravity theory whose flat limit is also non-degenerate.

Let us note that the Maxwell Carroll and the MEC algebras correspond to particular sub-cases of the infinite-dimensional Carrollian Maxwell algebra defined in \cite{Gomis:2019nih}. To our knowledge, the enlarged Carroll and the EEC algebras obtained here does not belong to a generalized family of UR infinite-dimensional algebras. 
It would be interesting to explore if, analogously to the Maxwellian case, the enlarged Carroll and its extended version can be seen as particular sub-cases of an infinite family of UR generalizations. In particular, one could expect that such generalization can be related to the UR expansion of the Maxwell algebra introduced in \cite{Gomis:2019nih} through a vanishing cosmological constant limit $\ell\rightarrow\infty$. 

On the other hand, as it has been shown that UR symmetry groups play a remarkable role in the contexts of flat holography (i.e., holography of flat space) and fluid/gravity correspondence \cite{Bagchi:2010zz, Bagchi:2012cy, Bagchi:2016bcd, Lodato:2016alv, Bagchi:2019xfx, Ciambelli:2018xat, Ciambelli:2018wre, Ciambelli:2018ojf, Campoleoni:2018ltl}, it would be intriguing to study the holographic implications of the MEC and EEC symmetries, in particular in connection with the fluid/gravity correspondence.
Furthermore, it would be worth it to study supersymmetric extensions of the UR algebras and CS theories developed in the present paper, following the lines of what was done in, e.g., \cite{Ravera:2019ize, Ali:2019jjp} in the case of AdS Carroll CS supergravity and flat limit. From the results we have obtained at the purely bosonic level, in the supersymmetric case we expect non-trivial bosonic and, in particular, fermionic enlargment and extensions to play a prominent role in the CS construction.

It would be interesting to go further in the study of the Maxwell and AdS-$\mathcal{L}$ symmetries. In particular, one could explore the complete schematic ``cube'' summarizing the sequential contractions starting from the Maxwell and AdS-$\mathcal{L}$ algebras. The NR version of both symmetries was recently introduced in \cite{Aviles:2018jzw,Concha:2019lhn}. In this paper, we have presented the respective UR counterparts with a detailed analysis of the non-degeneracy of the invariant tensor. It would be worth it to elucidate the respective Static Hopf and para-Galilei version of the Maxwell and AdS-$\mathcal{L}$ algebras.

\section*{Acknowledgments}

This work was funded by the National Agency for Research and Development ANID (ex-CONICYT) - PAI grant No. 77190078 (P.C.) and FONDECYT grants No. 1211077 (P.C.). This work was supported by the Research project Code DIREG$\_$09/2020 (P.C.) of the Universidad Católica de la Santisima Concepción, Chile. P.C. would like to thank to the Dirección de Investigación and Vice-rectoría de Investigación of the Universidad Católica de la Santísima Concepción, Chile, for their constant support. L.R. would like to thank the Department of Applied Science and Technology of the Polytechnic University of Turin, and in particular Laura Andrianopoli and Francesco Raffa, for financial support.


\bibliographystyle{fullsort.bst}
 
\bibliography{Maxwellian_Carroll}

\providecommand{\href}[2]{#2}\begingroup\raggedright\begin{thebibliography}{10}

\bibitem{LevyLeblond}
J.~Lévy-Leblond, ``{Une nouvelle limite non-relativiste du groupe de
  poincaré},'' {\em Annales de l'institut Henri Poincaré (A) Physique
  théorique} {\bf 3} (1965) 1--12.

\bibitem{Bacry:1968zf}
H.~Bacry and J.~Levy-Leblond, ``{Possible kinematics},'' {\em J. Math. Phys.}
  {\bf 9} (1968) 1605--1614.

\bibitem{Gibbons:2002tv}
G.~Gibbons, K.~Hashimoto, and P.~Yi, ``{Tachyon condensates, Carrollian
  contraction of Lorentz group, and fundamental strings},'' {\em JHEP} {\bf 09}
  (2002) 061, \href{http://www.arXiv.org/abs/hep-th/0209034}{{\tt
  hep-th/0209034}}.

\bibitem{Hofman:2014loa}
D.~M. Hofman and B.~Rollier, ``{Warped Conformal Field Theory as Lower Spin
  Gravity},'' {\em Nucl. Phys. B} {\bf 897} (2015) 1--38,
  \href{http://www.arXiv.org/abs/1411.0672}{{\tt 1411.0672}}.

\bibitem{Bagchi:2013bga}
A.~Bagchi, ``{Tensionless Strings and Galilean Conformal Algebra},'' {\em JHEP}
  {\bf 05} (2013) 141, \href{http://www.arXiv.org/abs/1303.0291}{{\tt
  1303.0291}}.

\bibitem{Bagchi:2015nca}
A.~Bagchi, S.~Chakrabortty, and P.~Parekh, ``{Tensionless Strings from
  Worldsheet Symmetries},'' {\em JHEP} {\bf 01} (2016) 158,
  \href{http://www.arXiv.org/abs/1507.04361}{{\tt 1507.04361}}.

\bibitem{Bagchi:2016yyf}
A.~Bagchi, S.~Chakrabortty, and P.~Parekh, ``{Tensionless Superstrings: View
  from the Worldsheet},'' {\em JHEP} {\bf 10} (2016) 113,
  \href{http://www.arXiv.org/abs/1606.09628}{{\tt 1606.09628}}.

\bibitem{Bagchi:2017cte}
A.~Bagchi, A.~Banerjee, S.~Chakrabortty, and P.~Parekh, ``{Inhomogeneous
  Tensionless Superstrings},'' {\em JHEP} {\bf 02} (2018) 065,
  \href{http://www.arXiv.org/abs/1710.03482}{{\tt 1710.03482}}.

\bibitem{Bagchi:2018wsn}
A.~Bagchi, A.~Banerjee, S.~Chakrabortty, and P.~Parekh, ``{Exotic Origins of
  Tensionless Superstrings},'' {\em Phys. Lett. B} {\bf 801} (2020) 135139,
  \href{http://www.arXiv.org/abs/1811.10877}{{\tt 1811.10877}}.

\bibitem{Duval:2014uoa}
C.~Duval, G.~W. Gibbons, P.~A. Horvathy, and P.~M. Zhang, ``{Carroll versus
  Newton and Galilei: two dual non-Einsteinian concepts of time},'' {\em Class.
  Quant. Grav.} {\bf 31} (2014) 085016,
  \href{http://www.arXiv.org/abs/1402.0657}{{\tt 1402.0657}}.

\bibitem{Clark:2016qbj}
T.~E. Clark and T.~ter Veldhuis, ``{AdS-Carroll Branes},'' {\em J. Math. Phys.}
  {\bf 57} (2016), no.~11, 112303,
  \href{http://www.arXiv.org/abs/1605.05484}{{\tt 1605.05484}}.

\bibitem{Basu:2018dub}
R.~Basu and U.~N. Chowdhury, ``{Dynamical structure of Carrollian
  Electrodynamics},'' {\em JHEP} {\bf 04} (2018) 111,
  \href{http://www.arXiv.org/abs/1802.09366}{{\tt 1802.09366}}.

\bibitem{Roychowdhury:2019aoi}
D.~Roychowdhury, ``{Carroll membranes},'' {\em JHEP} {\bf 10} (2019) 258,
  \href{http://www.arXiv.org/abs/1908.07280}{{\tt 1908.07280}}.

\bibitem{Bergshoeff:2015wma}
E.~Bergshoeff, J.~Gomis, and L.~Parra, ``{The Symmetries of the Carroll
  Superparticle},'' {\em J. Phys. A} {\bf 49} (2016), no.~18, 185402,
  \href{http://www.arXiv.org/abs/1503.06083}{{\tt 1503.06083}}.

\bibitem{Hartong:2015xda}
J.~Hartong, ``{Gauging the Carroll Algebra and Ultra-Relativistic Gravity},''
  {\em JHEP} {\bf 08} (2015) 069,
  \href{http://www.arXiv.org/abs/1505.05011}{{\tt 1505.05011}}.

\bibitem{Bergshoeff:2016soe}
E.~Bergshoeff, D.~Grumiller, S.~Prohazka, and J.~Rosseel, ``{Three-dimensional
  Spin-3 Theories Based on General Kinematical Algebras},'' {\em JHEP} {\bf 01}
  (2017) 114, \href{http://www.arXiv.org/abs/1612.02277}{{\tt 1612.02277}}.

\bibitem{Bergshoeff:2017btm}
E.~Bergshoeff, J.~Gomis, B.~Rollier, J.~Rosseel, and T.~ter Veldhuis,
  ``{Carroll versus Galilei Gravity},'' {\em JHEP} {\bf 03} (2017) 165,
  \href{http://www.arXiv.org/abs/1701.06156}{{\tt 1701.06156}}.

\bibitem{Matulich:2019cdo}
J.~Matulich, S.~Prohazka, and J.~Salzer, ``{Limits of three-dimensional gravity
  and metric kinematical Lie algebras in any dimension},'' {\em JHEP} {\bf 07}
  (2019) 118, \href{http://www.arXiv.org/abs/1903.09165}{{\tt 1903.09165}}.

\bibitem{Ravera:2019ize}
L.~Ravera, ``{AdS Carroll Chern-Simons supergravity in 2 + 1 dimensions and its
  flat limit},'' {\em Phys. Lett. B} {\bf 795} (2019) 331--338,
  \href{http://www.arXiv.org/abs/1905.00766}{{\tt 1905.00766}}.

\bibitem{Ali:2019jjp}
F.~Ali and L.~Ravera, ``{$\mathcal{N}$-extended Chern-Simons Carrollian
  supergravities in $2+1$ spacetime dimensions},'' {\em JHEP} {\bf 02} (2020)
  128, \href{http://www.arXiv.org/abs/1912.04172}{{\tt 1912.04172}}.

\bibitem{Grumiller:2020elf}
D.~Grumiller, J.~Hartong, S.~Prohazka, and J.~Salzer, ``{Limits of JT
  gravity},'' {\em JHEP} {\bf 02} (2021) 134,
  \href{http://www.arXiv.org/abs/2011.13870}{{\tt 2011.13870}}.

\bibitem{Gomis:2020wxp}
J.~Gomis, D.~Hidalgo, and P.~Salgado-Rebolledo, ``{Non-relativistic and
  Carrollian limits of Jackiw-Teitelboim gravity},'' {\em JHEP} {\bf 05} (2021)
  162, \href{http://www.arXiv.org/abs/2011.15053}{{\tt 2011.15053}}.

\bibitem{Schrader:1972zd}
R.~Schrader, ``{The maxwell group and the quantum theory of particles in
  classical homogeneous electromagnetic fields},'' {\em Fortsch. Phys.} {\bf
  20} (1972) 701--734.

\bibitem{Bacry:1970ye}
H.~Bacry, P.~Combe, and J.~Richard, ``{Group-theoretical analysis of elementary
  particles in an external electromagnetic field. 1. the relativistic particle
  in a constant and uniform field},'' {\em Nuovo Cim. A} {\bf 67} (1970)
  267--299.

\bibitem{Bacry:1970du}
H.~Bacry, P.~Combe, and J.~L. Richard, ``{Group-theoretical analysis of
  elementary particles in an external electromagnetic field. 2. the
  nonrelativistic particle in a constant and uniform field},'' {\em Nuovo Cim.
  A} {\bf 70} (1970) 289--312.

\bibitem{Gomis:2017cmt}
J.~Gomis and A.~Kleinschmidt, ``{On free Lie algebras and particles in
  electro-magnetic fields},'' {\em JHEP} {\bf 07} (2017) 085,
  \href{http://www.arXiv.org/abs/1705.05854}{{\tt 1705.05854}}.

\bibitem{Edelstein:2006se}
J.~D. Edelstein, M.~Hassaine, R.~Troncoso, and J.~Zanelli, ``{Lie-algebra
  expansions, Chern-Simons theories and the Einstein-Hilbert Lagrangian},''
  {\em Phys. Lett. B} {\bf 640} (2006) 278--284,
  \href{http://www.arXiv.org/abs/hep-th/0605174}{{\tt hep-th/0605174}}.

\bibitem{Izaurieta:2009hz}
F.~Izaurieta, E.~Rodriguez, P.~Minning, P.~Salgado, and A.~Perez, ``{Standard
  General Relativity from Chern-Simons Gravity},'' {\em Phys. Lett. B} {\bf
  678} (2009) 213--217, \href{http://www.arXiv.org/abs/0905.2187}{{\tt
  0905.2187}}.

\bibitem{Concha:2013uhq}
P.~Concha, D.~Pe\~nafiel, E.~Rodr\'\i{}guez, and P.~Salgado,
  ``{Even-dimensional General Relativity from Born-Infeld gravity},'' {\em
  Phys. Lett. B} {\bf 725} (2013) 419--424,
  \href{http://www.arXiv.org/abs/1309.0062}{{\tt 1309.0062}}.

\bibitem{Concha:2014vka}
P.~Concha, D.~Penafiel, E.~Rodriguez, and P.~Salgado, ``{Chern-Simons and
  Born-Infeld gravity theories and Maxwell algebras type},'' {\em Eur. Phys. J.
  C} {\bf 74} (2014) 2741, \href{http://www.arXiv.org/abs/1402.0023}{{\tt
  1402.0023}}.

\bibitem{Cangemi:1992ri}
D.~Cangemi, ``{One formulation for both lineal gravities through a dimensional
  reduction},'' {\em Phys. Lett. B} {\bf 297} (1992) 261--265,
  \href{http://www.arXiv.org/abs/gr-qc/9207004}{{\tt gr-qc/9207004}}.

\bibitem{Duval:2008tr}
C.~Duval, Z.~Horvath, and P.~A. Horvathy, ``{Chern-Simons gravity, based on a
  non-semisimple group},'' \href{http://www.arXiv.org/abs/0807.0977}{{\tt
  0807.0977}}.

\bibitem{Salgado:2014jka}
P.~Salgado, R.~J. Szabo, and O.~Valdivia, ``{Topological gravity and
  transgression holography},'' {\em Phys. Rev. D} {\bf 89} (2014), no.~8,
  084077, \href{http://www.arXiv.org/abs/1401.3653}{{\tt 1401.3653}}.

\bibitem{Hoseinzadeh:2014bla}
S.~Hoseinzadeh and A.~Rezaei-Aghdam, ``{(2$+$1)-dimensional gravity from
  Maxwell and semisimple extension of the Poincar\'e gauge symmetric models},''
  {\em Phys. Rev. D} {\bf 90} (2014), no.~8, 084008,
  \href{http://www.arXiv.org/abs/1402.0320}{{\tt 1402.0320}}.

\bibitem{Caroca:2017izc}
R.~Caroca, P.~Concha, O.~Fierro, E.~Rodr\'\i{}guez, and P.~Salgado-Rebolledo,
  ``{Generalized Chern\textendash{}Simons higher-spin gravity theories in three
  dimensions},'' {\em Nucl. Phys. B} {\bf 934} (2018) 240--264,
  \href{http://www.arXiv.org/abs/1712.09975}{{\tt 1712.09975}}.

\bibitem{Concha:2018zeb}
P.~Concha, N.~Merino, O.~Miskovic, E.~Rodr\'\i{}guez, P.~Salgado-Rebolledo, and
  O.~Valdivia, ``{Asymptotic symmetries of three-dimensional Chern-Simons
  gravity for the Maxwell algebra},'' {\em JHEP} {\bf 10} (2018) 079,
  \href{http://www.arXiv.org/abs/1805.08834}{{\tt 1805.08834}}.

\bibitem{Concha:2018jxx}
P.~Concha, D.~M. Peñafiel, and E.~Rodríguez, ``{On the Maxwell supergravity
  and flat limit in 2 + 1 dimensions},'' {\em Phys. Lett. B} {\bf 785} (2018)
  247--253, \href{http://www.arXiv.org/abs/1807.00194}{{\tt 1807.00194}}.

\bibitem{Concha:2019icz}
P.~Concha, ``{N-extended Maxwell supergravities as Chern-Simons theories in
  three spacetime dimensions},'' {\em Phys. Lett. B} {\bf 792} (2019) 290--297,
  \href{http://www.arXiv.org/abs/1903.03081}{{\tt 1903.03081}}.

\bibitem{Chernyavsky:2020fqs}
D.~Chernyavsky, N.~S. Deger, and D.~Sorokin, ``{Spontaneously broken $3d$
  Hietarinta/Maxwell Chern--Simons theory and minimal massive gravity},'' {\em
  Eur. Phys. J. C} {\bf 80} (2020), no.~6, 556,
  \href{http://www.arXiv.org/abs/2002.07592}{{\tt 2002.07592}}.

\bibitem{Adami:2020xkm}
H.~Adami, P.~Concha, E.~Rodriguez, and H.~Safari, ``{Asymptotic symmetries of
  Maxwell Chern\textendash{}Simons gravity with torsion},'' {\em Eur. Phys. J.
  C} {\bf 80} (2020), no.~10, 967,
  \href{http://www.arXiv.org/abs/2005.07690}{{\tt 2005.07690}}.

\bibitem{Caroca:2021bjo}
R.~Caroca, P.~Concha, J.~Matulich, E.~Rodr\'\i{}guez, and D.~Tempo,
  ``{Hypersymmetric extensions of Maxwell Chern-Simons gravity in (2+1)
  dimensions},'' \href{http://www.arXiv.org/abs/2105.12243}{{\tt 2105.12243}}.

\bibitem{deAzcarraga:2010sw}
J.~A. de~Azcarraga, K.~Kamimura, and J.~Lukierski, ``{Generalized cosmological
  term from Maxwell symmetries},'' {\em Phys. Rev. D} {\bf 83} (2011) 124036,
  \href{http://www.arXiv.org/abs/1012.4402}{{\tt 1012.4402}}.

\bibitem{Durka:2011nf}
R.~Durka, J.~Kowalski-Glikman, and M.~Szczachor, ``{Gauged AdS-Maxwell algebra
  and gravity},'' {\em Mod. Phys. Lett. A} {\bf 26} (2011) 2689--2696,
  \href{http://www.arXiv.org/abs/1107.4728}{{\tt 1107.4728}}.

\bibitem{deAzcarraga:2012qj}
J.~A. de~Azcarraga, K.~Kamimura, and J.~Lukierski, ``{Maxwell symmetries and
  some applications},'' {\em Int. J. Mod. Phys. Conf. Ser.} {\bf 23} (2013)
  01160, \href{http://www.arXiv.org/abs/1201.2850}{{\tt 1201.2850}}.

\bibitem{Concha:2014tca}
P.~Concha and E.~Rodr\'\i{}guez, ``{N = 1 Supergravity and Maxwell
  superalgebras},'' {\em JHEP} {\bf 09} (2014) 090,
  \href{http://www.arXiv.org/abs/1407.4635}{{\tt 1407.4635}}.

\bibitem{Penafiel:2017wfr}
D.~M. Pe\~nafiel and L.~Ravera, ``{On the Hidden Maxwell Superalgebra
  underlying D=4 Supergravity},'' {\em Fortsch. Phys.} {\bf 65} (2017), no.~9,
  1700005, \href{http://www.arXiv.org/abs/1701.04234}{{\tt 1701.04234}}.

\bibitem{Ravera:2018vra}
L.~Ravera, ``{Hidden role of Maxwell superalgebras in the free differential
  algebras of D = 4 and D = 11 supergravity},'' {\em Eur. Phys. J. C} {\bf 78}
  (2018), no.~3, 211, \href{http://www.arXiv.org/abs/1801.08860}{{\tt
  1801.08860}}.

\bibitem{Concha:2018ywv}
P.~Concha, L.~Ravera, and E.~Rodríguez, ``{On the supersymmetry invariance of
  flat supergravity with boundary},'' {\em JHEP} {\bf 01} (2019) 192,
  \href{http://www.arXiv.org/abs/1809.07871}{{\tt 1809.07871}}.

\bibitem{Salgado-Rebolledo:2019kft}
P.~Salgado-Rebolledo, ``{The Maxwell group in 2+1 dimensions and its
  infinite-dimensional enhancements},'' {\em JHEP} {\bf 10} (2019) 039,
  \href{http://www.arXiv.org/abs/1905.09421}{{\tt 1905.09421}}.

\bibitem{Concha:2019eip}
P.~Concha and H.~Safari, ``{On Stabilization of Maxwell-BMS Algebra},'' {\em
  JHEP} {\bf 04} (2020) 073, \href{http://www.arXiv.org/abs/1909.12827}{{\tt
  1909.12827}}.

\bibitem{Kibaroglu:2020tbr}
S.~Kibaroglu and O.~Cebecioglu, ``{Gauge theory of the Maxwell and semi-simple
  extended (anti) de Sitter algebra},''
  \href{http://www.arXiv.org/abs/2007.14795}{{\tt 2007.14795}}.

\bibitem{Cebecioglu:2021dqb}
O.~Cebecioglu and S.~Kibaroglu, ``{Maxwell-Modified Metric Affine Gravity},''
  \href{http://www.arXiv.org/abs/2104.12670}{{\tt 2104.12670}}.

\bibitem{Diaz:2012zza}
J.~Diaz, O.~Fierro, F.~Izaurieta, N.~Merino, E.~Rodriguez, P.~Salgado, and
  O.~Valdivia, ``{A generalized action for (2 + 1)-dimensional Chern-Simons
  gravity},'' {\em J. Phys. A} {\bf 45} (2012) 255207,
  \href{http://www.arXiv.org/abs/1311.2215}{{\tt 1311.2215}}.

\bibitem{Fierro:2014lka}
O.~Fierro, F.~Izaurieta, P.~Salgado, and O.~Valdivia, ``{Minimal AdS-Lorentz
  supergravity in three-dimensions},'' {\em Phys. Lett. B} {\bf 788} (2019)
  198--205, \href{http://www.arXiv.org/abs/1401.3697}{{\tt 1401.3697}}.

\bibitem{Concha:2018jjj}
P.~Concha, N.~Merino, E.~Rodríguez, P.~Salgado-Rebolledo, and O.~Valdivia,
  ``{Semi-simple enlargement of the $\mathfrak{bms}_3$ algebra from a
  $\mathfrak{so}(2,2)\oplus\mathfrak{so}(2,1)$ Chern-Simons theory},'' {\em
  JHEP} {\bf 02} (2019) 002, \href{http://www.arXiv.org/abs/1810.12256}{{\tt
  1810.12256}}.

\bibitem{Concha:2016kdz}
P.~Concha, R.~Durka, C.~Inostroza, N.~Merino, and E.~Rodríguez, ``{Pure
  Lovelock gravity and Chern-Simons theory},'' {\em Phys. Rev. D} {\bf 94}
  (2016), no.~2, 024055, \href{http://www.arXiv.org/abs/1603.09424}{{\tt
  1603.09424}}.

\bibitem{Concha:2017nca}
P.~Concha and E.~Rodríguez, ``{Generalized Pure Lovelock Gravity},'' {\em
  Phys. Lett. B} {\bf 774} (2017) 616--622,
  \href{http://www.arXiv.org/abs/1708.08827}{{\tt 1708.08827}}.

\bibitem{Ipinza:2016con}
M.~Ipinza, P.~Concha, L.~Ravera, and E.~Rodríguez, ``{On the Supersymmetric
  Extension of Gauss-Bonnet like Gravity},'' {\em JHEP} {\bf 09} (2016) 007,
  \href{http://www.arXiv.org/abs/1607.00373}{{\tt 1607.00373}}.

\bibitem{Penafiel:2018vpe}
D.~M. Peñafiel and L.~Ravera, ``{Generalized cosmological term in $D=4$
  supergravity from a new AdS--Lorentz superalgebra},'' {\em Eur. Phys. J. C}
  {\bf 78} (2018), no.~11, 945, \href{http://www.arXiv.org/abs/1807.07673}{{\tt
  1807.07673}}.

\bibitem{Banaudi:2018zmh}
A.~Banaudi and L.~Ravera, ``{Generalized AdS-Lorentz deformed supergravity on a
  manifold with boundary},'' {\em Eur. Phys. J. Plus} {\bf 133} (2018), no.~12,
  514, \href{http://www.arXiv.org/abs/1803.08738}{{\tt 1803.08738}}.

\bibitem{Bonanos:2008ez}
S.~Bonanos and J.~Gomis, ``{Infinite Sequence of Poincare Group Extensions:
  Structure and Dynamics},'' {\em J. Phys. A} {\bf 43} (2010) 015201,
  \href{http://www.arXiv.org/abs/0812.4140}{{\tt 0812.4140}}.

\bibitem{Bonanos:2008kr}
S.~Bonanos and J.~Gomis, ``{A Note on the Chevalley-Eilenberg Cohomology for
  the Galilei and Poincare Algebras},'' {\em J. Phys. A} {\bf 42} (2009)
  145206, \href{http://www.arXiv.org/abs/0808.2243}{{\tt 0808.2243}}.

\bibitem{Gomis:2009dm}
J.~Gomis, K.~Kamimura, and J.~Lukierski, ``{Deformations of Maxwell algebra and
  their Dynamical Realizations},'' {\em JHEP} {\bf 08} (2009) 039,
  \href{http://www.arXiv.org/abs/0906.4464}{{\tt 0906.4464}}.

\bibitem{Aviles:2018jzw}
L.~Avilés, E.~Frodden, J.~Gomis, D.~Hidalgo, and J.~Zanelli,
  ``{Non-Relativistic Maxwell Chern-Simons Gravity},'' {\em JHEP} {\bf 05}
  (2018) 047, \href{http://www.arXiv.org/abs/1802.08453}{{\tt 1802.08453}}.

\bibitem{Concha:2019lhn}
P.~Concha and E.~Rodríguez, ``{Non-Relativistic Gravity Theory based on an
  Enlargement of the Extended Bargmann Algebra},'' {\em JHEP} {\bf 07} (2019)
  085, \href{http://www.arXiv.org/abs/1906.00086}{{\tt 1906.00086}}.

\bibitem{Penafiel:2019czp}
D.~M. Peñafiel and P.~Salgado-Rebolledo, ``{Non-relativistic symmetries in
  three space-time dimensions and the Nappi-Witten algebra},'' {\em Phys. Lett.
  B} {\bf 798} (2019) 135005, \href{http://www.arXiv.org/abs/1906.02161}{{\tt
  1906.02161}}.

\bibitem{Concha:2020sjt}
P.~Concha, M.~Ipinza, and E.~Rodríguez, ``{Generalized Maxwellian exotic
  Bargmann gravity theory in three spacetime dimensions},'' {\em Phys. Lett. B}
  {\bf 807} (2020) 135593, \href{http://www.arXiv.org/abs/2004.01203}{{\tt
  2004.01203}}.

\bibitem{Concha:2020ebl}
P.~Concha, L.~Ravera, E.~Rodr\'\i{}guez, and G.~Rubio, ``{Three-dimensional
  Maxwellian Extended Newtonian gravity and flat limit},'' {\em JHEP} {\bf 10}
  (2020) 181, \href{http://www.arXiv.org/abs/2006.13128}{{\tt 2006.13128}}.

\bibitem{Concha:2019mxx}
P.~Concha, L.~Ravera, and E.~Rodríguez, ``{Three-dimensional Maxwellian
  extended Bargmann supergravity},'' {\em JHEP} {\bf 04} (2020) 051,
  \href{http://www.arXiv.org/abs/1912.09477}{{\tt 1912.09477}}.

\bibitem{Concha:2020eam}
P.~Concha, M.~Ipinza, L.~Ravera, and E.~Rodr\'\i{}guez, ``{Non-relativistic
  three-dimensional supergravity theories and semigroup expansion method},''
  \href{http://www.arXiv.org/abs/2010.01216}{{\tt 2010.01216}}.

\bibitem{Gomis:2019nih}
J.~Gomis, A.~Kleinschmidt, J.~Palmkvist, and P.~Salgado-Rebolledo,
  ``{Newton-Hooke/Carrollian expansions of (A)dS and Chern-Simons gravity},''
  {\em JHEP} {\bf 02} (2020) 009,
  \href{http://www.arXiv.org/abs/1912.07564}{{\tt 1912.07564}}.

\bibitem{Soroka:2006aj}
D.~V. Soroka and V.~A. Soroka, ``{Semi-simple extension of the (super)Poincare
  algebra},'' {\em Adv. High Energy Phys.} {\bf 2009} (2009) 234147,
  \href{http://www.arXiv.org/abs/hep-th/0605251}{{\tt hep-th/0605251}}.

\bibitem{Witten:1988hc}
E.~Witten, ``{(2+1)-Dimensional Gravity as an Exactly Soluble System},'' {\em
  Nucl. Phys. B} {\bf 311} (1988) 46.

\bibitem{Izaurieta:2006zz}
F.~Izaurieta, E.~Rodriguez, and P.~Salgado, ``{Expanding Lie (super)algebras
  through Abelian semigroups},'' {\em J. Math. Phys.} {\bf 47} (2006) 123512,
  \href{http://www.arXiv.org/abs/hep-th/0606215}{{\tt hep-th/0606215}}.

\bibitem{RUBIO:2020mnh}
G.~Rubio, {\em {Gravedad newtoniana en teor\'\i{}as de gauge extendidas}}.
\newblock PhD thesis, Concepcion U., 11, 2020.

\bibitem{Bagchi:2010zz}
A.~Bagchi, ``{Correspondence between Asymptotically Flat Spacetimes and
  Nonrelativistic Conformal Field Theories},'' {\em Phys. Rev. Lett.} {\bf 105}
  (2010) 171601, \href{http://www.arXiv.org/abs/1006.3354}{{\tt 1006.3354}}.

\bibitem{Bagchi:2012cy}
A.~Bagchi and R.~Fareghbal, ``{BMS/GCA Redux: Towards Flatspace Holography from
  Non-Relativistic Symmetries},'' {\em JHEP} {\bf 10} (2012) 092,
  \href{http://www.arXiv.org/abs/1203.5795}{{\tt 1203.5795}}.

\bibitem{Bagchi:2016bcd}
A.~Bagchi, R.~Basu, A.~Kakkar, and A.~Mehra, ``{Flat Holography: Aspects of the
  dual field theory},'' {\em JHEP} {\bf 12} (2016) 147,
  \href{http://www.arXiv.org/abs/1609.06203}{{\tt 1609.06203}}.

\bibitem{Lodato:2016alv}
I.~Lodato and W.~Merbis, ``{Super-BMS$_{3}$ algebras from $ \mathcal{N}=2 $
  flat supergravities},'' {\em JHEP} {\bf 11} (2016) 150,
  \href{http://www.arXiv.org/abs/1610.07506}{{\tt 1610.07506}}.

\bibitem{Bagchi:2019xfx}
A.~Bagchi, A.~Mehra, and P.~Nandi, ``{Field Theories with Conformal Carrollian
  Symmetry},'' {\em JHEP} {\bf 05} (2019) 108,
  \href{http://www.arXiv.org/abs/1901.10147}{{\tt 1901.10147}}.

\bibitem{Ciambelli:2018xat}
L.~Ciambelli, C.~Marteau, A.~C. Petkou, P.~M. Petropoulos, and K.~Siampos,
  ``{Covariant Galilean versus Carrollian hydrodynamics from relativistic
  fluids},'' {\em Class. Quant. Grav.} {\bf 35} (2018), no.~16, 165001,
  \href{http://www.arXiv.org/abs/1802.05286}{{\tt 1802.05286}}.

\bibitem{Ciambelli:2018wre}
L.~Ciambelli, C.~Marteau, A.~C. Petkou, P.~M. Petropoulos, and K.~Siampos,
  ``{Flat holography and Carrollian fluids},'' {\em JHEP} {\bf 07} (2018) 165,
  \href{http://www.arXiv.org/abs/1802.06809}{{\tt 1802.06809}}.

\bibitem{Ciambelli:2018ojf}
L.~Ciambelli and C.~Marteau, ``{Carrollian conservation laws and Ricci-flat
  gravity},'' {\em Class. Quant. Grav.} {\bf 36} (2019), no.~8, 085004,
  \href{http://www.arXiv.org/abs/1810.11037}{{\tt 1810.11037}}.

\bibitem{Campoleoni:2018ltl}
A.~Campoleoni, L.~Ciambelli, C.~Marteau, P.~M. Petropoulos, and K.~Siampos,
  ``{Two-dimensional fluids and their holographic duals},'' {\em Nucl. Phys. B}
  {\bf 946} (2019) 114692, \href{http://www.arXiv.org/abs/1812.04019}{{\tt
  1812.04019}}.

\end{thebibliography}\endgroup

\end{document}